# Understanding Functional Protein-Protein Interactions of ABCB11 and ADA in Human and Mouse

Antara Sengupta
Dept. of MCA
MCKV Institute of Engineering
Howrah, India
Antara.sngpt@gmail.com

Sk. Sarif Hassan
Dept. of Mathematics
University of Petroleum and Energy Studies
Dehradun, India
sarif.hassan@icts.res.in

Pabitra Pal Choudhury
Applied Statistics Unit
Indian Statistical Institute
Kolkata, India
pabitrapalchoudhury@gmail.com

*Abstract*— **Proteins are macromolecules which hardly act alone; they need to make interactions with some other proteins to do so. Numerous factors are there which can regulate the interactions between proteins [4]. Here in this present study we aim to understand Protein -Protein Interactions (PPIs) of two proteins ABCB11 and ADA from quantitative point of view. One of our major aims also is to study the factors that regulate the PPIs and thus to distinguish these PPIs with proper quantification across the two species Homo Sapiens and Mus Musculus respectively to know how one protein interacts with different set of proteins in different species.**

## Introduction

Proteins are the workhouses where most of the biological processes in a cell take place. Proteins are macromolecules and need associations of some other proteins to participate in various essential molecular processes within a cell and there is a talk about Protein- Protein Interactions (PPIs) [7].Numerous factors are there which can regulate the PPIs. Abnormal PPIs can lead to develop the basis of diseases like cancers. Not only that but drugs also target PPIs for *interfacial inhibition*. As an example Brefeldin A (BFA) [1] acts as inhibitor and attacks macromolecular complexes when the complex is in a transition state(structurally and energetically unbalanced) and goes for drug binding[3].Certain biological properties are there in gene as well as protein level which play significant role in transcriptions, protein structure formations, and gene expressions and can have significant role in protein network formation.

Moreover, proteins fold spontaneously into complicated three-dimensional structures which are essential for several biological activities. Mainly the driving energy for this folding process comes from the hydrophobic effect, Van der Waals forces, and salt bridges at specific binding domains on each protein [6]. The strength of the binding depends on the size of the binding domain. Leucine zipper is a common surface domain that can provide stable protein-protein interactions. It consists of α-helices on each protein that bind to each other in a parallel fashion through the hydrophobic bonding of regularly-spaced leucine residues on each α-helix that project between the adjacent helix peptide chains. Because of the tight molecular packing, leucine zippers provide stable binding for multi-protein complexes [7].

Quantitative understanding of genes/ proteins refers to its unique characterizations or a numeric vector corresponding to each sequence which would act as their signature by finding some quantitative attributes. The quantification essentially captures how the nucleotides or amino acids are arranged in the sequence [9]. Here in this paper it is tried to make quantitative analysis of the genes ABCB11 and ADA and the genes which are functionally attached with them across the two species *Homo Sapiens* and *Mus Musculus*. The analysis has been made with genes and Proteins at their sequence levels and as well as the physical properties of those protein to study the factors that regulate the PPIs and thus to distinguish these PPIs with proper quantification across the two species to know how one protein interacts with different set of proteins in different species.

A. Model Representation

We code any given DNA sequence to numeric sequence for further analysis. There are various ways of doing so. Here we have coded as A=1, T=2, G=3 and C=4 and thus a DNA sequence would be transformed to a sequence of corresponding numeric sequence.

B. Data Set Specification

In this paper as the PPIs of the proteins ABCB11 and ADA for the species Sapiens and *Mus Musculus* are taken from String Database which is shown below. It is to be noted that the associations of the proteins are purely functional associations. It can be observed that the proteins ABCB11 and ADA both have some proteins in the networks which are common for both the species, whereas some proteins are specifically participating in the specified species not in other.

TABLE I ABCB11 and ADA PPIs. Tick defines presence of the protein the species as mentioned.

| Proteins in the Network of ABCB11 | Homo Sapiens | *Mus Musculus* | Proteins in the Network of ADA | Homo Sapiens | *Mus Musculus* |
|---|---|---|---|---|---|
| Abcb11 | √ | √ | Ada | √ | √ |
| Alb | √ | √ | ADK | √ | √ |
| ATP8B1 | √ | | ADORA1 | √ | |
| Baat | √ | √ | DCK | √ | √ |
| Nr0b2 | √ | √ | DPP4 | √ | |
| NR5A2 | √ | √ | MYB | √ | |
| NRIH4 | √ | √ | NT5C | √ | √ |
| Slco10a1 | √ | √ | NT5C1 | √ | |
| Slco1a1 | | √ | Nt5c1a | | √ |
| SLCO1A2 | √ | | NT5C2 | √ | √ |
| Slco1a4 | | √ | Nt5c3 | | √ |
| Slco1a6 | | √ | NT5E | √ | √ |
| SLCO1B1 | √ | | Nt5m | | √ |
| Slco1b2 | | √ | PNP | √ | √ |
| SLCO1B3 | √ | | Pnp2 | | √ |

## II. METHODOLOGY

Factors that regulate the PPIs are known qualitatively. To model the regulatory factors mathematically it is necessary to characterize the PPIs in DNA and Protein sequence levels which are performed as follows:-

A. Quantitative understanding of DNA sequence

Quantitative understanding of a gene in molecular level can be done by applying several mathematical parameters [13][9] which can derive some data and can verify some fact. In this paper, to build quantitative model it is tried to find out the underlying geometries of DNA structure and the hidden geometrical rules, that is to capture spatial ordering of bases across the DNA and Protein so that we can make a mapping between those geometrical rules with biological activities of those proteins.

Quantitative understanding of DNA sequence is taken place through following phases:-

1) Generating Indicator Matrix and Calculating Fractal Dimension of the Indicator Matrices

In Mathematics the term 'Indicator' is a numerical measure of a quality or characteristic of some aspect of a program; evidence that something is occurring, that progress is being made. The notion of indicator matrix and its characterization through fractal dimension was proposed by Carlo Cattani [13][2]. DNA sequences have four basic components ($A$ = adenine, $C$ = cytosine, $G$ = guanine, $T$ = thymine) which is defined as four alphabets A, C, G, T respectively.

Let us consider $D \stackrel{\text{def}}{=} \{A, C, G, T\}$ be the set of nucleotides and $x \in D$ where, x is any alphabet of D. A DNA sequence is the finite symbolic string $S = \mathbb{N} \times D$, so that $S \stackrel{\text{def}}{=} \{x_h\}_{h=1,...,N}, N<\infty$ being $x_h \stackrel{\text{def}}{=} (h, x)=x(h)$ where $h=1,2,...,N$ and $x \in D$ the value $x$ at the position $h$. According to C.Cattani the Indicator Matrix can be characterized through fractal dimension as follows,

$$f: S \times S \to \{0,1\} \text{ such that } f(x_h, x_k) \stackrel{\text{def}}{=} \begin{cases} 1 & if x_h = x_k \\ 0 & if x_h \neq x_k \end{cases}$$

where $x_h, x_k \in S$

So now it is possible to easily describe a N×N sparse binary matrix from an indicator matrix of N length, which may be written as,

$$M_{hk} = f(x_k)x_h \; ; \; x_k \in S, h,k = 1,2,3,...N$$

But it is not possible to differentiate between zeros formed by distinct base pairs. So slight modification is needed into it, which is given below [7],

$$f: S \times S \to \{1,2,3,4\} \text{ such that}$$

$$(x_h, x_k) \stackrel{\text{def}}{=} \begin{cases} 1 & if x_h = x_k \\ 2 & if x_h \neq x_k \; ;x_1 \\ 3 & if x_h \neq x_k \; ;x_h \\ 4 & if x_h \neq x_k; x_h \end{cases}$$

So the matrix $M_{hk}$ will be decomposed into four binary matrices namely, A1, A2, A3, A4 and corresponding four binary images will be made for each DNA sequence.

The fractal dimensions of those indicator matrices have been calculated using "Box-Counting" method [17]. Box-Counting is a method of collection of data for analyzing complex pattern like dataset, images etc. and breaking them into smaller part or pieces and analyze them individually. Fractal dimensions for each indicator matrix are derived using BENOIT software.

Indicator matrices for ABCB11 and ADA and the genes in their networks are made for both the species. Fractal Dimensions (FDs) of four indicator matrices are also calculated for each species which are shown in Supplementary file 1and Supplementary file 2 respectively.

TABLE II. Interval of Fractal Dimensions and the names of genes which have similar characteristics to ABCB11 and ADA respectively

| Species | Gene Name | A1 | A2 | A3 | A4 | Genes having similar characteristics |
|---|---|---|---|---|---|---|
| Homo Sapiens | ABCB11 and genes in network | (1.79542, 1.87851) | (1.79782, 1.87844) | (1.78368, 1.87196) | (1.78112, 1.87203) | SLCO1A2 |
| Mus Musculus | | (1.82043, 1.89552) | (1.81957, 1.89538) | (1.81734, 1.88822) | (1.8167, 1.88831) | Alb |
| Homo Sapiens | ADA and genes in network | (1.79542, 1.87851) | (1.79782, 1.87844) | (1.78368, 1.87196) | (1.78112, 1.87203) | NT5C |
| Mus Musculus | | (1.81957, 1.89538) | (1.81957, 1.89538) | (1.81734, 1.88822) | (1.8222, 1.88831) | Nt5c3 |

1. The gene 'SLCO1A2'and 'alb' have similar characteristics of Abcb11 of *Homo Sapiens* and *Mus Musculus* respectively for FDs of all four indicator matrices.
2. The gene NT5C and Nt5c3 have similar characteristics of ADA of *Homo Sapiens* and *Mus Musculus* respectively for FDs of all four indicator matrices.

2) Fractal Dimension of DNA Walk

As defined in the [9][13],We can define DNA walk as,
$$a_n \stackrel{\text{def}}{=} \sum_{i=1}^{n} f(A, x_i), t_n \stackrel{\text{def}}{=} \sum_{i=1}^{n} f(T, x_i), g_n \stackrel{\text{def}}{=} \sum_{i=1}^{n} f(C, x_i), c_n \stackrel{\text{def}}{=} \sum_{i=1}^{n} f(G, x_i).$$

It has been resulted by plotting $(W_n, V_n)$ as two functions $W_n \stackrel{\text{def}}{=} \sin a_n^2 - \sin g_n^2$ and $V_n \sin t_n^2 - \sin c_n^2$ are defi

The Fractal Dimensions of all the genes are deduced using Box Counting method.

TABLE III. Interval of Fractal Dimensions of DNA Walk and the names of genes which have similar characteristics to ABCB11 and ADA respectively

| Species | Gene Name | FD of DNA WALK | Genes having similar Characteristics |
|---|---|---|---|
| Homo Sapiens | ABCB11 and genes in network | (1.78266, 1.90321) | BAAT |
| Mus Musculus | | (1.79281, 1.88461) | Slco1a1 |
| Homo Sapiens | ADA and genes in network | (1.78487, 1.94016) | PNP |
| Mus Musculus | | (1.79281, 1.88461) | Nt5e |

The DNA walk for all the genes in the PPI of ABC11 and ADA respectively are computed and then the fractal dimensions of the DNA are enumerated using Benoit. The detail result is given in Supplementary file 1 and 2.

1. The gene 'BAAT' and 'Slco1a1' have similar characteristics of Abcb11 of *Homo Sapiens* and *Mus Musculus* respectively for FDs of all four indicator matrices.
2. The gene PNP and Nt5e have similar characteristics of ADA of *Homo Sapiens* and *Mus Musculus* respectively for FDs of all four indicator matrices.

3) Poly-string Mean and Standard Deviation:-

A gene is a string constituting of different permutations of the base pairs A, C, T and G where repetition of a base pair is allowed. It is possible to classify the sequences based on the ordering of poly-string mean of A, C, T, and G in the string [12][13].

$$\text{Mean} N_u = \frac{2(N_{u1} + N_{u2} + N_{u3} + \cdots N_{um})}{m} \cdot (m+1) \quad (1)$$

Where, $N_{ui} \in \{A, T, C, G\}$, i=1,2,3,…,m and m is the length of longest poly-string over the string.

The poly-string mean and SD are calculated for all the genes of the PPI of ABCB11 and ADA respectively and the detail result is given in the supplementary files 1 and 2.

TABLE IV. The poly-string mean and SD for all the genes of the PPI of ABCB11 and ADA

| Species | Gene Name | Classifications based on poly-string mean | Classifications based on Poly-string SD |
|---|---|---|---|
| Homo Sapiens | ABCB11 and genes in network | 5 | 5 |
| Mus Musculus | | 7 | 8 |
| Homo Sapiens | ADA and genes in network | 6 | 8 |
| Mus Musculus | | 7 | 7 |

There are 256 possible poly-strings of length four using four nucleotides A, T, C, G. It can be observed that the genes in PPIs of ABCB11 for *Homo Sapiens* and *Mus Musculus* have been classified into 5 and 7 classes respectively according to poly-string mean whereas, the genes in PPIs of ADA have been classified into 5 and 8 classes respectively.

In the case of poly-string SD the genes in the PPIs of ABCB11 for *Homo Sapiens* and *Mus Musculus* have been classified into 6 and 7 classes respectively. The genes in the PPIs of ADA for

*Homo Sapiens* and *Mus Musculus* have been classified into 8 and 7 classes respectively.

B. Quantification of Bio-physical Properties

Followings are the properties of DNA sequences which are calculated by using Oligo Calc. These properties include the physical properties of oligonucelotides:-
  1) Physical Constants
        i.   Length
        ii.  Molecular weight
        iii. GC Contents
  2) Melting Temperature (TM) Calculations
        i.   Basic
        ii.  Salt Adjusted
  3) Thermodynamic Contents Conditions : 1M NaCl at 25°C at pH 7
        i.   Rlnk
        ii.  DeltaH
        iii. DeltaG
        iv.  DeltaS

The Bio-physical Properties are calculated for all the genes of the PPI of ABCB11 and ADA respectively are derived and the detail result is given in the supplementary files 1 and 2.

1. Overall observation for the species of *Homo sapiens* speaks that although the quantitative difference of the physical properties of the genes ABCB11 and SLCOA2 are small but GC count of Physical constants and Melting Temperature Calculations are same of ABCB11 and SLCO1B1 are same.
2. From the result it is observed for the species of *Homo sapiens* that majorly the quantitative differences of physical constants and Thermodynamic Contents Conditions of the genes ADA and NT5C2 are smaller but in the case of GC count and Melting Temperature (TM) Calculations the genes ADA and ADORA1 are having very less differences.
3. From the result it is observed for the species of *Mus Musculus* that quantitative differences of physical constants and Thermodynamic Contents Conditions of the genes ADA and NT5C2 are smallest and gene abcb11 have similarity of physical properties with Nr5a2.
4. The result as a whole enlightens that, physical properties of both of the species can't be equal.

C. Quantification of Protein Structural Properties

Here, we extend our quantitative analysis to the protein structural level and which is given below.
The protein properties which are considered here are viz. Alpha Helix (Chou Fasman), Beta Strand (Chou Fasman), Parallel Beta Strand, Anti Parallel Beta Strand, Molecular Beta Strand , Total Beta Strand, Coil, Hydrophobicity (Aboderin) and Polarity Grantham. The plotting of protein properties has been made using MATLAB (Bioinformatics Tools) and the fractal dimension has been deducted using BENOIT™ software.

TABLE IV. Ranges of protein properties of each species

| Species | *Homo Sapiens* | *Mus Musculus* | *Homo Sapiens* | *Mus Musculus* |
|---|---|---|---|---|
| Protein Name | ABCB11 & proteins in network | | ADA & proteins in network | |
| Alpha Helix | (1.93681, 1.94316) | (1.93031, 1.94357) | (1.93681, 1.94316) | (1.93957, 1.94348) |
| Antiparallel beta strand | (1.9361, 1.94326) | (1.93087, 1.94293) | (1.9361, 1.94326) | (1.93944, 1.94329) |
| Beta Sheet | (1.93688, 1.94354) | (1.93061, 1.94296) | (1.93688, 1.94354) | (1.93942, 1.94352) |
| Coil | (1.93972, 1.94426) | (1.93398, 1.94385) | (1.93972, 1.94426) | (1.94115, 1.94422) |
| Hydrophobicity | (1.93548, 1.94321 ) | (1.92922, 1.94312) | (1.93548, 1.94321) | (1.93924, 1.94322) |
| Molecular Beta Strand | (1.93906, 1.94411) | (1.93416, 1.94378) | (1.93906, 1.94411) | (1.93989, 1.9443) |
| Parallel Beta Strand | (1.93717, 1.94366) | (1.92921, 1.94295) | (1.93717, 1.94366) | (1.93832, 1.94368) |
| Polarity Grantham | (1.93577, 1.94368) | (1.93072, 1.94332) | (1.93577, 1.94368) | (1.94009, 1.94406) |
| Total Beta Strand | (1.9368, 1.94338) | (1.94271, 1.94273) | (1.9368, 1.94338) | (1.93908, 1.94345) |

From the result we can get the range of FDs within which the Protein Properties of the proteins ABCB11 and ADA and the proteins in their networks are lying for the 2 species.

The Protein Properties of ABCB11 are more similar with the properties of the proteins ATP8B1 for the species of *Homo Sapiens* but in case of *Mus Musculus* it has similarity with 'alb'. Whereas, it is worth enough to say that for both the species the Protein Properties of ADA are more similar with the properties of the protein ADK except the properties of Molecular Beta Strand and Parallel Beta Strand, which are more similar with PNP. But as a whole the quantitative understanding of the protein properties reveal that the protein PNP can have a continuous mapping with the gene ADA.

D. Quantitative comparisons of DNAs in terms of 2D graphical representations based on triplets

We know that Codon consists of three adjacent nucleotides which codes for a particular amino acid. Thus 64 codons are there which codes for 20 amino acids and 3 codons are there which cause termination of protein synthesis. Now each codons have their corresponding triplets of DNA which we can get from the DNA sequences. If we define a mapping '$\Psi$' to map each triplet to different weight, for any two pairs of triplets $(X_1, Y_1)$ and $(X_2, Y_2)$, where $X_1, Y_1, X_2, Y_2$, all are triplets, if the corresponding codons of $X_1$ and $Y_1$ code for the same amino acid but corresponding codons of $X_2$ and $Y_2$ code for different amino acid, thae that satisfy the following rule:

$$|\Psi(X_1) - \Psi(Y_1)| < |\Psi(X_2) - \Psi(Y_2)|$$

The mapping has been illustrated below in the tables V and VI:-

TABLE V. Codons and their corresponding Amino Acids.

| Codon | Corresponding Amino Acid | Codon | Corresponding Amino Acid |
|---|---|---|---|
| GCU, GCC, GCA, GCG | Alanine | UGG | Tryptophan |
| CUU, CUC, CUA, CUG, UUA, UUG | Leucine | CAU, CAC | Histidine |
| CGU, CGC, CGA, CGG, AGA, AGG | Arginine | UAU, UAC | Tyrosine |
| AAA, AAG | Lysine | AUU, AUC, AUA | Isoleucine |
| GAU, GAC | Aspartic acid | GUU, GUC, GUA, GUG | Valine |
| AUG | Methionine | UAA, UAG, UGA | Stop Codons |
| AAU, AAC | Asparagine | GAA, GAG | Glutamic acid |
| UUU, UUC | Phenylalanine | UCU, UCC, UCA, UCG, AGU,GC | Serine |
| UGU, UGC | Cysteine | CAA, CAG | Glutamine |
| CCU, CCC, CCA, CCG | Proline | ACU, ACC, ACA, ACG | Threonine |
| | | GGU, GGC, GGA, GGG | Glycine |

TABLE VI. Triplets and their corresponding weight ($\Psi$).

| Codon | Corresponding Triplet | Weight ($\Psi$) | Codon | Corresponding Triplet | Weight ($\Psi$) |
|---|---|---|---|---|---|
| GCU | GCT | 1.1 | CUU | CTT | 11.1 |
| GCC | GCC | 1.2 | CUC | CTC | 11.2 |
| GCA | GCA | 1.3 | CUA | CTA | 11.3 |
| GCG | GCG | 1.4 | CUG | CTG | 11.4 |
| CGU | CGT | 2.1 | UUA | TTA | 11.5 |
| CGC | CGC | 2.2 | UUG | TTG | 11.6 |
| CGA | CGA | 2.3 | AAA | AAA | 12.3 |
| CGG | CGG | 2.4 | AAG | AAG | 12.4 |
| AGA | AGA | 2.5 | UUU | TTT | 13.1 |
| AGG | AGG | 2.6 | UUC | TTC | 13.2 |
| GAU | GAT | 3.3 | CCU | CCT | 14.1 |
| GAC | GAC | 3.4 | CCC | CCC | 14.2 |
| AAU | AAT | 4.1 | CCA | CCA | 14.3 |
| AAC | AAC | 4.2 | CCG | CCG | 14.4 |
| UGU | TGT | 5.1 | UCU | TCT | 15.1 |
| UGC | TGC | 5.2 | UCC | TCC | 15.2 |
| GAA | GAA | 6.1 | UCA | TCA | 15.3 |
| GAG | GAG | 6.2 | UCG | TCG | 15.4 |
| CAA | CAA | 7.1 | AGU | AGT | 15.5 |
| CAG | CAG | 7.2 | AGC | AGC | 15.6 |
| GGU | GGT | 8.1 | ACU | ACT | 16.1 |
| GGC | GGC | 8.2 | ACC | ACC | 16.2 |
| GGA | GGA | 8.3 | ACA | ACA | 16.3 |
| GGG | GGG | 8.4 | ACG | ACG | 16.4 |
| CAU | CAT | 9.1 | UGG | TGG | 17.3 |
| CAC | CAC | 9.2 | UAU | TAT | 18.1 |
| AUU | ATT | 10.1 | UAC | TAC | 18.2 |
| AUC | ATC | 10.2 | GUU | GTT | 19.1 |
| AUA | ATA | 10.3 | GUC | GTC | 19.2 |
| UAA | TAA | 21.1 | GUA | GTA | 19.3 |
| UAG | TAG | 21.2 | GUG | GTG | 19.4 |
| UGA | TGA | 21.3 | AUG | ATG | 20.1 |

Let G=$g_1, g_2, g_3,……g_n$ is a DNA primary sequence, where $g_i$={A, T, C, G} for any i= 1, 2,…..n, so that, G=$t_1, t_2,……t_m$ be the corresponding triplet, where m=[n/3]. Then to plot G into a plot set, we can define mapping Θ as follows[18],

$$\Theta(G)=\{(1,\Psi(t_1)),(2,\Psi(t_2)),…..(m,(\Psi(t_m))\} \quad (2)$$

PPI Networks of ABCB11 and ADA for both the species reflects that some proteins are there, which are present in the networks of both species, like in case of ABCB11 (Abcb11,Alb,Baat, Nr0b2,NR5A2,NRIH4,Slco10a1) are common, whereas in case of ADA (ADA,ADK,DCK,NT5C,NT5C2,NT5E,PNP) are common. In this paper all DNA sequences of those genes are represented through 2D graph as graphical representation of DNA sequence provides viewing, sorting and even comparing various gene structures in simpler way. Thus quantitative comparisons of those DNAs are made possible in terms of 2D graphical representations.

One example of quantitative comparisons between DNAs of ABCB11 of *Homo Sapiens and Mus Musculus* in terms of 2D graph is shown below in figure 1:-

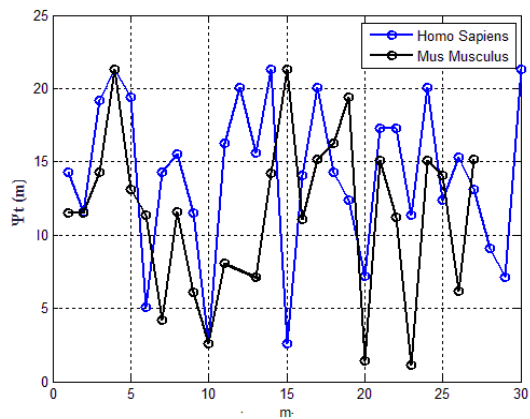

Fig 1: Quantitative comparisons of ABCB11 of *Homo sapiens and Mus musculus* of first 30 triplets

### III. CONCLUSION

In this paper, quantitative understanding and analysis of Functional Protein-Protein Interactions of ABCB11 and ADA in Human and Mouse has been taken place. Characterization and comparison of functionally associated PPIs through the light of different parameters have shown a statistical view of those factors which ultimately are responsible for network constructions of those two proteins for each species. Moreover, as PPI network of same protein may defer from species to species, the paper reflects a quantitative comparative view of PPI networks of same proteins but for different two species

### IV. APPENDIX

A. Supplementary file1 – *HOMO SAPIENS* DB.xls

B. Supplementary file2 – *MUS MUSCULUS*.xls